\documentclass[12pt,notoc]{JHEP3}

\usepackage{amsmath,amssymb,euscript,array}
\newcommand{\startappendix}{
\setcounter{section}{0}
\renewcommand{\thesection}{\Alph{section}}}
\newcommand{\Appendix}[1]{
\refstepcounter{section}
\begin{flushleft}
{\large\bf Appendix \thesection: #1}
\end{flushleft}}
\usepackage{epsfig}
\def\N{{\cal N}}

\def\Tr{{\rm Tr}}

\def\det{{\rm det}}

\newcommand{\ads}[1]{{\rm AdS}_{#1}}
\newcommand{\sph}[1]{{ S}^{#1}}

\def\Dbarslash{\,\,{\raise.15ex\hbox{/}\mkern-12mu {\bar D}}}
\def\Dslash{\,\,{\raise.15ex\hbox{/}\mkern-12mu D}}
\def\delslash{\,\,{\raise.15ex\hbox{/}\mkern-9mu \partial}}
\def\delbarslash{\,\,{\raise.15ex\hbox{/}\mkern-9mu {\bar\partial}}}

\newcommand{\EQ}[1]{\begin{equation} #1 \end{equation}}

\newcommand{\SP}[1]{\begin{equation}\begin{split} #1
\end{split}\end{equation}}

\title{Instabilities of the Small Black Hole: a view from ${\mathbf{{\cal N}=4}}$ SYM}
\author{Timothy Hollowood, S. Prem Kumar and Asad Naqvi \\
Department of Physics,\\ University of Wales Swansea,\\
Swansea, SA2 8PP, UK.\\
E-mail: {\tt t.hollowood,s.p.kumar,a.naqvi@swan.ac.uk}}
\preprint{hep-th/0607111 \\ SWAT/06/471}
\abstract{We compute a one-loop effective action for the constant
  modes of the scalars and the Polyakov loop matrix of $\N=4$ SYM on
  $\sph{3}$ at finite temperature and weak 't Hooft coupling. Above a
  critical temperature, the effective potential develops new
  unstable directions accompanied by new saddle points which only
  preserve an
  $SO(5)$ subgroup of the $SO(6)$ global R-symmetry. We
  identify this phenomenon as the weak coupling version of the
  well known Gregory-Laflamme localization instability in the gravity
  dual of the strongly 
   coupled field theory: The small $\ads{5}$ black hole when viewed as a ten
  dimensional, asymptotically  $\ads{5}\times \sph{5}$ solution
  smeared on the $\sph{5}$ is unstable to localization on $\sph{5}$.  
  Our effective potential, in a specific Lorentzian continuation, can
  provide a qualitative holographic description of the decay of
  the ``topological black hole'' into the AdS bubble of nothing.

} 

\begin{document}
\section{Introduction}
The finite temperature behaviour of $SU(N)$ gauge theories at large
$N$ provides a remarkable window into the physics of black holes and
stringy gravity via the AdS/CFT correspondence
\cite{Maldacena:1997re}. Features 
of the phase structure of large $N$, $\N=4$ supersymmetric Yang-Mills (SYM)
theory on $\sph{3}$ at finite temperature are now known to mirror aspects of
semiclassical gravity on asymptotically $\ads{5}$ spacetimes
\cite{Witten:1998qj,Witten:1998zw,
  Aharony:1999ti,Sundborg:1999ue,Aharony:2003sx}.   
Importantly, the work of \cite{Aharony:2003sx} has
demonstrated that studying the field theory in the tractable regime of
{\em weak}  't Hooft coupling may allow us to deduce qualitative
physics of the string theory dual emerging at intermediate and strong
couplings. The 
weakly coupled field theory exists in one of two thermodynamically stable
phases separated by a first order deconfinement transition; in
addition there is the possibility of a thermodynamically unstable
saddle point \cite{Aharony:2003sx, Aharony:2005bq,
  Alvarez-Gaume:2005fv}. Indeed, the 
strong coupling 
gravity dual also exhibits a  
first order Hawking-Page transition 
\cite{Hawking:1982dh}
 between two stable geometries: thermal AdS space
and the big AdS-Schwarzschild black hole, mediated by the
thermodynamically unstable Euclidean small AdS black hole bounce.
\footnote{The comparison between the phase structure of
  the weakly interacting thermal gauge theory and strong coupling
  gravity dual has recently been extended to include non-zero
  chemical potentials for the global $U(1)^3\subset SO(6)_R$ charges in the
  field theory \cite{Yamada:2006rx} (see also \cite{Harmark:2006di}). The latter is dual to charged
  black hole geometries in 
  $\ads{5}$.} 

The aim of this paper is to take the qualitative matching of
thermodynamic phase structure one step further and to explore certain
questions which have dynamical consequences. Specifically, the small
$\ads{5}$ Schwarzschild black hole, when viewed as a ten dimensional
asymptotically $\ads{5} \times \sph{5}$ solution  smeared uniformly on
the $\sph{5}$, has a classical dynamical instability (in addition to
having a thermodynamic instability). This is a Gregory-Laflamme 
instability to localization
\cite{Gregory:1993vy,Gubser:2000ec,Gubser:2000mm, Reall:2001ag,
  Hubeny:2002xn}, wherein the small black hole originally smeared on the
$\sph{5}$ develops an instability as its horizon size in $\ads{5}$
decreases below 
a critical radius (or equivalently, above a critical temperature). The
unstable   
mode leads to the small black hole becoming non-uniform and eventually
point-like on the $\sph{5}$, breaking the associated $SO(6)$
isometry down to $SO(5)$. 
The question we aim to answer in
this paper
is how, if at all, this dynamical instability to localization may be
seen within the framework of the holographically dual thermal field
theory on a three-sphere. Remarkably, within the regime of validity of
perturbation theory   
we find that above a critical temperature at which the radius of
the three-sphere becomes comparable to the Debye screening
length, the weakly coupled
field theory exhibits a clear
signal of such an instability.\footnote{A similar phenomenon has been
  studied in 1+1 dimensional Yang-Mills theory in
  \cite{Aharony:2004ig}.}

The $\N=4$ theory on $\sph{3}$ at finite temperature has two
 scales, namely the radius $R$ of the three-sphere and the
temperature $T$. Consequently the theory possesses two tunable dimensionless
parameters, the combination $(TR)$ and the 't Hooft coupling
$\lambda=g^2_{YM} N$. On $\sph{3}\times\sph{1}$, at weak 't Hooft
coupling $\lambda\ll1$,  we calculate a one
loop quantum effective action as a function of homogeneous background
expectation values for the Polyakov loop 
\footnote{We use the term Polyakov loop loosely since we are actually
 referring to the holonomy matrix of the gauge field around the
 thermal circle and not just its trace.} 
and the six scalar fields
transforming in the adjoint representation of $SU(N)$. 
More precisely, we compute a finite temperature
effective action of the Coleman-Weinberg type on a slice of the full
configuration space parameterized by the $N$ eigenvalues of each of
these fields. The one loop computation is valid for a wide range
of temperatures at weak 't Hooft coupling 
\EQ{0\leq\;TR\;\ll {1\over \lambda}~.}
A crucial point, which ensures the validity of perturbation theory at
high temperatures within the range above, is the infrared cutoff
provided by the finite size of the three-sphere. In particular, we
 will mainly be interested in temperatures  $(TR )\sim 1/\sqrt\lambda
 $ which is well within the above range. At these temperatures  the
 radius of $\sph{3}$ becomes 
comparable to the Debye screening length $(\sqrt\lambda T)^{-1}$. In
flat space, at these scales one needs to resum the one loop thermal mass of the
modes to cure infrared divergences in perturbation theory. On
$\sph{3}$ however, the presence of an explicit infrared cutoff, namely
$R\sim (\sqrt\lambda T)^{-1}$, ensures the validity of perturbation
theory at these length scales, albeit in the parameter $\sqrt\lambda$
as opposed to $\lambda$. Our calculation differs from the
work of \cite{Yamada:2006rx} in that we have generic values for the
 Polyakov loop matrix and scalar
 fields, but vanishing chemical potential. Thus we explore the
 landscape of the effective potential away 
 from local minima.

We now summarize the main features of the quantum effective action
we obtain in equations  
\eqref{sbosonic}, \eqref{sfermionic} and \eqref{full} of the
paper. Firstly, the   
effective action provides a combined static 
effective potential on the space of the eigenvalues of the Polyakov
loop matrix and of the constant modes of the scalar fields in the $\N=4$
multiplet. It is obtained  
by integrating out all fluctuations of these matrices (including
off-diagonal ones) and
all non-zero modes on $\sph{3}\times \sph{1}$.
For vanishing scalar fields, our action reduces to the unitary matrix
model obtained in \cite{Aharony:2003sx}. Hence at low temperatures $T\ll
R^{-1}$, the effective potential exhibits a single saddle point with
vanishing scalar fields and uniformly distributed Polyakov loop
eigenvalues. This is the ``thermal AdS'' saddle point. At the Hagedorn
temperature $T_H= -( R\;\ln(7-4\sqrt 3))^{-1}$ (or slightly below it 
for finite coupling) a first order
deconfinement transition occurs beyond which a new global minimum
appears with a gapped distribution for the Polyakov loop eigenvalues
and vanishing scalar fields. This is the ``big AdS black hole''
minimum. 

At temperatures above $T_H$, the thermal AdS saddle point with
vanishing scalar fields persists as a thermodynamically
unstable saddle point, while the globally stable big black hole
dominates the canonical ensemble. For all 
temperatures $T_H < T < R^{-1}/\sqrt\lambda$, this continues to be the
picture with the scalar fields forced to vanish
by the tree level quadratic mass term arising from the conformal
coupling to the curvature of $\sph{3}$.

At temperatures $T\gtrsim R^{-1}/\sqrt \lambda$ however, something
interesting begins to happen. The one loop contribution becomes
comparable to the tree level mass while, crucially, the higher loop
corrections remain perturbatively small, suppressed by powers
of $\sqrt\lambda$. At a critical temperature $T_c\sim R^{-1}/\sqrt
\lambda$, a new unstable mode, along the R-charged scalar directions, 
emerges at the thermodynamically
unstable ``thermal AdS'' saddle point. In the large $N$ limit this
critical temperature is found to be
\EQ{T_c={\pi\over \sqrt {2\lambda}}\;R^{-1}.}
Beyond this temperature, new unstable saddle points
appear with lower action than the ``thermal AdS''
configuration which becomes unstable to rolling down to these new
extrema. Importantly, at these extrema the scalar fields of the $\N=4$
theory have non-zero values. In the large $N$ theory these expectation
values are left invariant only  by an $SO(5)$ subgroup of the full 
$SO(6)$ R-symmetry of
the theory. 
However,
the ``big black hole'' configuration continues to be the global
minimum of the action with vanishing VEVs for all
scalars which is consistent with expectations from the bulk gravitational physics. 

 \begin{figure}[ht] 
\centerline{\includegraphics[width=3in]{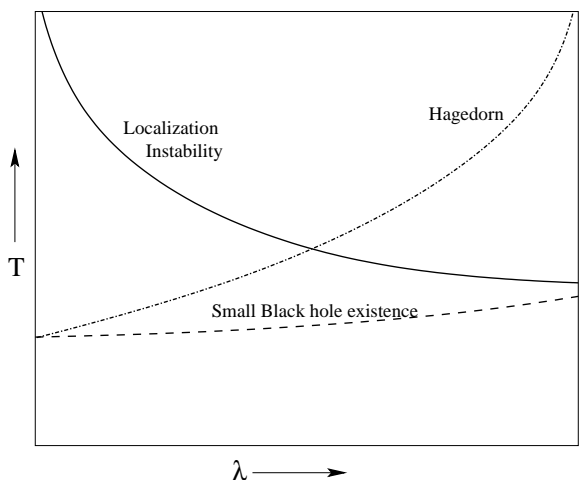}}
\caption{\footnotesize Qualitative plot of small black hole (SBH) existence
  lines, and extrapolation of weak coupling critical temperature for
  the Gregory-Laflamme localization instability as a function of 't
  Hooft coupling 
  $\lambda$. SBH comes into existence above the dashed line 
and below the Hagedorn temperature line. Localization occurs above the
  solid line.}  
\label{phase}
\end{figure}

We identify this high temperature phenomenon
as a continuation to weak 't Hooft coupling of the   
Gregory-Laflamme instability encountered in the gravity
dual of the strongly coupled gauge theory. There too, a
thermodynamically unstable saddle point, namely the small AdS black
hole develops a new dynamical instability to localization on the
$\sph{5}$ breaking the $SO(6)$ isometry to $SO(5)$. The main
difference is that at weak coupling, at temperatures of
$O({R^{-1}/\sqrt\lambda})$ any ``small black hole'' type unstable
saddle point disappears \cite{Aharony:2003sx} or merges with the
``thermal AdS'' configuration \cite{Alvarez-Gaume:2005fv}.

In fact, as the temperature is increased, we find more and more
unstable modes in the large $N$ field theory around the thermal AdS
saddle point. Whenever the temperature hits the critical values 
\EQ{
T_c^{(l)}= {(2l-1) \pi  \over {\sqrt{2 \lambda}}},}
labeled by the positive integers $l$, a new unstable mode
appears. This phenomenon is consistent with the observations of
\cite{Hubeny:2002xn} in the context of the instabilities of the small
black hole where, as the temperature is increased, new tachyonic modes emerge.
Importantly, the unstable modes involve only the homogeneous
fields on $S^3$, all Kaluza-Klein harmonics remain massive.
 
A naive extrapolation of our
weak coupling result suggests that the critical temperature for the 
onset of the instability decreases with increasing $\lambda$. It is
therefore plausible that for large couplings, the localization
instability kicks in at a temperature where the small black hole
saddle point still exists and one obtains a 
Gregory-Laflamme instability. This is illustrated in Figure \ref{phase}.
We must
also bear in mind that weak 't 
Hooft coupling translates to string scale curvatures in the dual
geometry and at these scales it is possible that  
the small black hole makes a Horowitz-Polchinski transition to a
highly excited state of strings \cite{Horowitz:1996nw,
  Alvarez-Gaume:2005fv, Alvarez-Gaume:2006jg} before the onset of any
localization instabilities.

We should point out that the large $N$ limit plays an
important role in all of 
the above, since only in this limit can we legitimately speak of 
``$SO(6)$ breaking expectation values'' in the field theory formulated
on the compact 
space. On a compact space, in the quantum theory we must integrate
over all points in field space which are related by the global
symmetry. \footnote{ For a discussion of these issues particularly from
  the viewpoint of the small black hole gravity solution, see
  \cite{Alvarez-Gaume:2006jg}.} At large $N$
the integration measure for this averaging
procedure yields a contribution which is sub-leading in $N$. We discuss
this in more detail later in Section 3.3.2.

The outline of our paper is as follows. In Section 2 we review the
general story of black hole instabilities in AdS space. Section 3 is
devoted to a detailed calculation of the one loop effective potential
in field theory and
determining its saddle points as a function of temperature. We
present our conclusions, interpretations and questions for future
study in Section 4. The analysis of unstable directions of the
effective potential is presented in an Appendix.

\section{Instabilities and AdS Schwarzschild Black Holes}
We begin by reviewing the physics of Schwarzschild black holes in
asymptotically AdS spaces \cite{Hawking:1982dh,
  Witten:1998zw,Witten:1998qj}. Of particular interest is the 
the five dimensional AdS-Schwarzschild black hole which is an asymptotically
$\ads{5}$  solution to the vacuum Einstein's equation with a negative
cosmological constant: 
\EQ{
ds^2= - V(r) dt^2 + {dr^2 \over V(r)} + r^2 d\Omega_3^2,
}
where $V(r)= 1+{r^2 \over R^2} - {r_h^2  \over r^2}(1+{r_h \over
  R^2})$ and $R$ is the radius of $\ads{5}$. The black hole horizon 
is at $r=r_h$ where $V(r)$ vanishes. This black hole emits
Hawking radiation. To find the associated Hawking temperature at which
the black hole is 
in equilibrium with a thermal bath at that temperature, we employ the usual
trick of going to the Euclidean section and requiring that there are
no conical singularities. Taking $ t \rightarrow -i \chi$, the
resulting Euclidean metric is 
\EQ{ ds^2 = V(r) d\chi^2 + {dr^2 \over V(r)} + r^2 d \Omega_3^2.}
Requiring that the metric has no conical singularity at $r=r_h$ leads
to the $\chi$ coordinate being periodic with period $\beta=  {2\pi R^2
r_h \over 2r_h^2+R^2}$. 
%
Asymptotically, as $r \rightarrow \infty$, the space is Euclidean
$\ads{5}$ with the Euclidean time direction being a circle of  
 circumference $\beta$. This implies
 an equilibrium Hawking temperature  
 \EQ{
 T= {1 \over \beta}= {2 r_h^2 +R^2 \over 2\pi r_h R^2}.
 }
Notice that for a certain range of $T$, there are two solutions for $r_h$ given by:
\EQ{
r_h = { \pi T R^2  \over 2} \Bigl( 1 \pm  \sqrt{1- {2 \over (\pi RT)^2}} \Bigr).
}
 \begin{figure}[ht] 
\centerline{\includegraphics[width=2in]{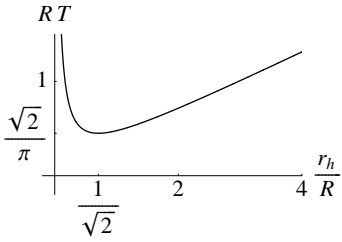}}
\caption{\footnotesize Temperature as a function of the horizon radius.}  
\label{temp}
\end{figure}
There are no solutions unless $T  \geq {\sqrt{2} \over \pi R}\equiv
T_0$, which implies that the AdS Schwarzschild black holes only exist
for temperatures above this critical temperature. When $T > T_0$,
there are two geometries corresponding to the two values of $r_h$, the
small and the large AdS black holes. The specific heat $C_v \sim {dT
  \over dr_h }= {2r_h^2 -R^2 \over 2\pi R^2 r_h^2}$ is negative for
the small black hole and positive for the large black hole. This
implies that the small black hole is in an unstable equilibrium with
the radiation bath. At equilibrium, the Hawking radiation
emitted and 
the radiation absorbed from the heat bath are equal. If the black hole
emits a little more than it absorbs, it decreases in size, thereby
increasing its temperature (because of the negative specific heat)
which means it emits even more, implying the existence of a
thermodynamic instability to decay to thermal AdS. If on the other
hand, the small black hole emits infinitesimally less than it absorbs,
it will grow in size until it becomes the large black hole.  The large
black hole is in stable thermal equilibrium by virtue of its positive
specific heat.  In the Euclidean setup, the thermodynamic instability
of the small black hole manifests itself in the existence of a
non-conformal negative mode in the small fluctuation analysis.

Although the small $\ads{5}$ Schwarzschild black hole is
thermodynamically unstable, it suffers from no classical dynamical
instability, in the sense that there are no tachyonic small
fluctuation modes. This changes when we add extra dimensions along
which the black hole is uniformly smeared. For example, in the context
of type IIB string theory on asymptotically $\ads{5} \times \sph{5}$
background, the five dimensional small AdS black hole solution is
smeared uniformly on the $\sph{5}$. In what follows, we will review
how this 10 dimensional solution suffers from a classical dynamical
instability  to localization on the $\sph{5}$ \cite{Hubeny:2002xn}. 

\subsection{Gregory-Laflamme instabilities and the Gubser-Mitra
  conjecture}

When black holes are smeared along some extra non
compact ``internal'' direction,  
often  the onset of  a thermodynamic instability is accompanied by a
classical dynamical instability signaled by the
existence of tachyonic modes in a Lorentzian small fluctuation
analysis. In fact, this dynamical instability is of a Gregory-Laflamme
type \cite{Gregory:1993vy}: The instability is to localization in
the extra directions. 
We will now review  this in more detail.  
Such a link between thermodynamical
and dynamical instabilities was conjectured by Gubser and Mitra  
\cite{Gubser:2000ec,Gubser:2000mm}. Further evidence for this
connection was provided by Reall  
\cite{Reall:2001ag} and we briefly review his argument by considering
a black string in a five dimensional asymptotically flat space  \cite{Hubeny:2002xn}.  
 
 The black string in five dimensions is translationally invariant in
 one spatial direction. It can be viewed as a four dimensional
 Schwarzschild black hole solution smeared uniformly along one extra
 non-compact direction.  
If we consider just the four dimensional black hole in the Euclidean
section, the Euclidean Lichnerowicz operator $\Delta^{(4)}_E$ has a
negative eigen-mode 
\EQ{
\Delta^{(4)}_E h_{\mu \nu}= - \eta^2 h_{\mu \nu},
\label{lichnerowicz}
}
where $\eta$ is a real. The association of this Euclidean negative
mode with the thermodynamic instability  signaled by a negative
specific heat was made in \cite{Gregory:2001bd}. If we consider {\em static}
fluctuations in Lorentzian signature, the equations obtained are the
same as the Euclidean ones: $\Delta_{L}^{(4)} = \Delta_E^{(4)}$ when
acting on static fluctuations.

However, (\ref{lichnerowicz}) is not
the relevant equation in the Lorentzian space. The linearized equation governing
the small fluctuations about a Lorentzian 4D black hole solution is 
\EQ{\Delta^{(4)}_L h_{\mu \nu}=0.}
If, on the other
hand, we are considering a five-dimensional black string and
fluctuations which carry momentum along the translationally invariant
fifth direction, $h^{(5)}_{\mu \nu}(x,z)= h^{(4)}_{\mu \nu}(x)
e^{ikz}$, the relevant fluctuation equation involves a
five-dimensional Lichnerowicz operator:    
\EQ{
\Delta^{(5)}_L h^{(5)}_{\mu \nu} =  (\Delta^{(4)}_L h^{(4)}_{\mu \nu} +
k^2 h_{\mu \nu}^{(4)})e^{ikz}=0 . 
}  
This is the same as (\ref{lichnerowicz}) with $k^2= \eta^2$.  For this
value of $k$, there is a  ``modulus'' corresponding to this mode,
which signals the onset of instability in Lorentzian signature. This
was called the {\it threshold unstable mode} in \cite{Reall:2001ag}.   

The above analysis may seem to imply that every small AdS black hole,
when viewed as  solution in $\ads{5} \times \sph{5}$ which is
smeared on the $\sph{5}$ should exhibit a Gregory-Laflamme like
instability. However, the situation is more subtle if the internal
smeared directions are compact, as was discussed in
\cite{Hubeny:2002xn}. This is simply because the momentum in the compact
internal dimensions is quantized and this leads
to a constraint on the negative Euclidean eigenvalue for the existence
of a threshold unstable mode. For the smeared $\ads{5}\times \sph{5}$
Schwarzschild case the negative eigenvalue has 
to be ${-l(l+4) \over R^2}$ which are the eigenvalues of the Laplacian on the
$\sph{5}$. This in
turn imposes a constraint on the horizon radius of the
Gregory-Laflamme unstable black holes. For example, only for $r_h =
0.4 R$ does the $l=1$ mode on the $\sph{5}$ becomes unstable
\cite{Hubeny:2002xn}. This is smaller than $r_h = {R \over \sqrt{2}}$
which is the largest possible horizon radius for the small black hole.  In
other words, although the small black hole exists for $T \geq T_0
={\sqrt{2} \over \pi R}= {0.45 \over R}$, it only becomes
Gregory-Laflamme unstable to localization on the $\sph{5}$ for $T \geq
{0.53 \over R}$.  

\subsection{Hawking-Page transition and the big AdS Schwarzschild
  black hole} 

The AdS/CFT correspondence relates the bulk string theory partition
function on asymptotically $\ads{5} \times \sph{5}$ geometries to the
partition function of the boundary SYM theory. In the semiclassical
supergravity limit, the bulk partition function gets contributions
from saddle points which are classical solutions to the equations of
motion. Depending on the boundary geometry, there can be more than one
bulk saddle points and in such a situation a careful sum 
\cite{Witten:1998qj,Witten:1998zw} over all the
saddle points is required. \footnote{For a review of AdS/CFT
  correspondence with an emphasis over the sum over different bulk
  saddle points, see \cite{deBoer:2004yu}.}

For example, the
appropriate boundary geometry to calculate the canonical ensemble
partition function is $\sph{3} \times \sph{1}$. For temperatures $T >
T_0 = {\sqrt{2} \over \pi R}$, there are  three bulk saddle
points with this boundary behavior. These are the two AdS
Schwarzschild black holes (the small and the large) and the 
thermal AdS geometry, which is Euclidean AdS with a periodic time
direction. For temperatures lower than the critical temperature $T_0$,
thermal AdS is the only saddle point. The different saddle
points correspond to different phases, and the locally stable saddle
point with the  
least action dominates the canonical ensemble. 

The actions of the
three saddle points are formally infinite because of the infinite volume of
thermal AdS and AdS Schwarzschild black holes.  
 \begin{figure}[ht] 
\centerline{\includegraphics[width=2in]{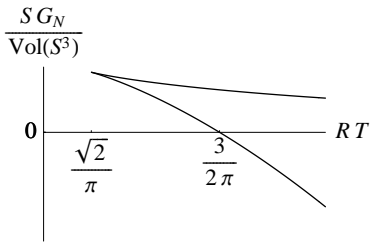}}
\caption{\footnotesize The actions for the big and the small black
  holes as a function of temperatures.}   
\label{actions}
\end{figure}
However, it turns out that the differences between the 
actions of the different solutions
are finite and 
subtracting the thermal AdS action from the action of the AdS
Schwarzschild black holes \cite{Hawking:1982dh} yields finite
results. Taking the thermal AdS action to be zero, the action for the 
AdS Schwarzschild black holes is given by  
\EQ{S_{\rm BH}={{\rm Vol}(S^3)r_h^3  \over 8 G_N}{R^2 -r_h^2 \over  (
    R^2+2 r_h^2  )}.} 
It is easy to see that the action of the small black hole is always
greater than both the action of the large black hole and that of
thermal AdS (Figure \ref{actions}), so the small black hole never
dominates the canonical ensemble. The action of the large black hole
becomes less than that of thermal AdS at the Hawking-Page transition
temperature $T_{HP}={3 \over 2 \pi
  R}$. At lower temperatures, thermal AdS dominates  while at higher
temperatures the large black hole dominates the canonical
ensemble.

\section{Weakly-Coupled ${\mathbf{{\cal N}=4}}$ SYM at Finite
  Temperature}

Recently, weakly coupled $SU(N)$ Yang-Mills theories on $\sph{3}$ have
been studied for large $N$ at finite temperature
\cite{Aharony:2003sx,Aharony:2005bq}, where it was shown that these
theories exhibit 
an interesting phase structure at weak coupling. The order
parameter distinguishing these phases is the expectation value of the
Polyakov loop, which is 
non-zero in the high temperature phase and is zero in the low
temperature phase. 

In fact, the theories have an interesting phase
structure even in the limit of zero coupling  \cite{Sundborg:1999ue},
\cite{Aharony:2003sx}. This is because if we  define the free
theory as 
a limit of the interacting theory, we need to impose the  
Gauss's law constraint
which enforces the zero charge condition on a spatially compact
manifold. On a finite $\sph{3}$, this effectively introduces
interactions which are non-trivial enough to lead to a first order 
Hagedorn type deconfinement 
phase transition at high temperatures. This Hagedorn phase
transition is believed to be the zero coupling analog of the
Hawking-Page transition in the dual geometry at strong coupling. At
non-zero but weak 't Hooft couplings it is expected that there will be
two phase transitions, a deconfining transition followed by the
Hagedorn phase transition.

In \cite{Aharony:2003sx}, the above phenomena at zero coupling were 
deduced from 
a Wilsonian effective action for the Polyakov loop matrix obtained by 
integrating out all non-zero modes on $\sph{3}\times\sph{1}$. The
effective action for the Polyakov loop matrix turns out to be a unitary
matrix model. In the large-$N$ limit it is natural to introduce a
continuous spectral density function $\rho(\alpha)$ for  
eigenvalues of the Polyakov loop, which are distributed on the unit
circle in the complex plane. It was shown in \cite{Aharony:2003sx}
that in the low temperature phase, the uniform distribution of
eigenvalues of the Polyakov loop is a stationary point of the
effective action and in fact is the minimum. The uniform distribution
signals a confined phase of the field theory and is the weak
coupling continuation of the thermal AdS dual geometry.

In the high temperature
phase, although the uniform distribution remains a saddle point, it is
not a minimum. The temperature at which the uniform distribution
saddle point ceases to be a minimum is the Hagedorn temperature, which
corresponds to the temperature of the confinement/deconfinement phase
transition in the free field limit.  
Above the Hagedorn temperature, the minimum action eigenvalue
distribution becomes nonuniform and in particular, is gapped, {\em
  i.e.} the spectral density $\rho(\alpha)$ necessarily 
vanishes on a subset of the circle. This implies a finite free energy
for colored external sources and is naturally associated to a
deconfined phase. This is the weak coupling version of the big AdS
black hole.

The small AdS black hole unstable saddle point cannot, however, be seen in the
free field limit. It is expected to exist at weak, non-zero 't
Hooft coupling \cite{Aharony:2003sx,Aharony:2005bq, Alvarez-Gaume:2005fv}. 

\subsection{Gregory-Laflamme in field theory}

Our goal in this section is to see the analog of the Gregory-Laflamme
instability 
in the weakly coupled ${\cal N}=4$ SYM at finite temperature on an $\sph{3}$. 
As we discussed in the previous section, this instability is
associated with the localization of the smeared small black hole on 
the 
$\sph{5}$, breaking the associated $SO(6)$ isometry. 
In the dual field theory we would expect this phenomenon to manifest
itself as an unstable saddle point which breaks 
the $SO(6)$ R-symmetry. Such symmetry
breaking has to be a subtle dynamical phenomenon for two reasons: i) 
The tree-level scalar potential on
$\sph{3}\times\sph{1}$ precludes a non-zero classical VEV for the
scalar 
fields in the ${\cal N}=4$ multiplet; ii)
field theories on compact spaces do not usually exhibit  
spontaneous symmetry breaking. 
 
To see the above phenomenon in weakly coupled ${\cal N}=4$ SYM 
we need to compute a quantum
effective potential for the spatial zero modes of the scalar fields
which 
are charged under the $SO(6)$ R-symmetry. Thus in addition to the
Polyakov loop, we will turn on the zero momentum modes of the six
scalar 
fields and obtain a joint effective  action for these degrees of 
freedom in perturbation theory. In addition, we will also see that
spontaneous breaking of the global R-symmetry can occur on a compact 
space, due to the large $N$ limit.

Finally, it is important to note that we will compute the quantum 
effective potential along a special subspace in the space of field
configurations, namely one 
where all the scalar fields and the Polyakov loop are simultaneously 
diagonalizable. This will be sufficient to explore the onset of the
localization phenomenon which we are interested in.

\subsubsection{The quantum effective potential}
In the canonical ensemble, the thermal partition function of a quantum
field 
theory is equal to the Euclidean path integral of the theory with a
periodic time direction of period $\beta = {1 \over T}$ with
anti-periodic boundary conditions for the fermions along the
temporal circle.  Since we are considering the ${\cal N}=4$
SYM on $\sph{3}$ at finite temperature, we perform a Euclidean path
integral for the theory on $\sph{3} \times \sph{1}$. 

We now describe
the calculation of this path integral and hence the effective free
energy (which is just log of the partition function) to one loop
order. This will be a Euclidean 1PI-effective action for static,
spatially homogeneous fields and will therefore be interpreted as a
static effective potential. We note here that although, strictly
speaking, the static  
effective potential is a useful tool for identifying the equilibrium
and locally stable field configurations, we will actually be
interested in exploring its features away from local minima.

Since the ${\cal N}=4$ theory is conformally invariant, the
ratio of the radii of the $\sph{3}$ and the
$\sph{1}$ is the only physically relevant dimensionless parameter (in
addition to the 't Hooft coupling constant)
\EQ{{R_{\sph{3}}\over R_{\sph{1}}}={R \over\beta}=R\;T,}
where the radius of the thermal circle is $\beta$, the inverse
temperature and that of the $\sph{3}$ is $R$. 

To do the effective potential calculation,  it is necessary to expand
all fields onto  
modes corresponding to the spherical harmonics on the $\sph{3}$ and
Fourier modes on the $\sph{1}$. We then compute an  
effective potential for the constant modes on the $\sph{3} \times
\sph{1}$ by integrating out all other modes in the non-trivial
background of the constant modes. 
In particular, we will compute the effective potential for the modes  
$\alpha$  and $\varphi_a$,
\EQ{\alpha \; = \;{T\over {\rm Vol}(\sph{3})}\int_{\sph{3}\times
    \sph{1}} A_0,}
and 
\EQ{\varphi_a \; = \;{T\over {\rm Vol}(\sph{3})}\int_{\sph{3}\times
    \sph{1}} \phi_a\;\;,~~~~~{a=1 \cdots 6}.} 
These are the  spatially homogeneous, time independent pieces
of $A_0$ 
(the gauge field component along $\sph{1}$) and
$\phi_a$, 
respectively, and so are constant on $\sph{3}\times \sph{1}$.

Note that at the classical level, the Wilson line $\alpha$ is a true zero mode 
in the sense that the quadratic tree level action is independent of it.
In contrast, the scalar modes $\varphi_a$ are not zero modes of the
classical theory since conformally coupled
scalars have a mass of the order of the inverse radius of $\sph{3}$,
due to the curvature of the three-sphere. 
\\\\
{\underline{\it Tree level potential:}}\\\\
The action for the bosonic fields is
\EQ{
S_{\rm b}=\frac1{g^2_{YM}}\int
d^4x\sqrt{g}\,\text{Tr}\,\Big\{\tfrac14F_{\mu\nu}F^{\mu\nu}
+\tfrac12D_\mu\phi_a D^\mu\phi_a+\tfrac1{2R^2}\phi_a\phi_a-
\tfrac14[\phi_a,\phi_b]^2\Big\}\ .
}
Here, summation over $SO(6)$ indices is implied.
The mass term above arises from the conformal coupling of the scalar
fields to the curvature of $\sph{3}$ and in our conventions this mass is
$R^{-2}$.

In order to integrate out the
non-zero momentum modes, we first shift the scalar fields by a
background homogeneous mode
\EQ{
\phi_a\to\varphi_a+\phi_a\;.} 
The resulting theory has the following tree
level effective action for the scalar modes $\varphi_a$: 
\EQ{
S^{(0)} ={\beta R^3 \pi^2}{1\over g^2_{YM}}\;\Tr\left(-[\alpha,\varphi_a]^2-
{1\over 2}[\varphi_a,\varphi_b]^2+\,\varphi_a^2\;R^{-2} \right).\label{tree}
}
In flat space in the absence of curvature induced mass terms, this
scalar potential would have led to a Coulomb branch of vacua parametrized by
the eigenvalues of a mutually commuting set of $\alpha$ and
$\varphi_a$. However, the conformal mass term for the $\{\varphi_a\}$
forces these to vanish at tree level since the potential \eqref{tree} is a
sum of positive definite quantities. In addition to the above 
terms there will be radiative corrections at finite temperature,
obtained by integrating out the matter and gauge field fluctuations at
one loop level. The nature of these radiative corrections and the
question of stability of the $\varphi_a=0$ classical solution will be 
the subject of the calculation below.
\\\\
{\underline{\it Radiative correction at one loop:}}
\\\\
Computing the quantum corrections to the classical potential for 
generic homogeneous background fields is technically
difficult. Instead,  we
will compute the quantum corrected potential for mutually
commuting background field configurations, namely those satisfying
\EQ{[\alpha,\varphi_a]=[\varphi_a,\varphi_b]=0.}
This will allow us to explore only a slice of the full configuration
space. However it will be sufficient to address the issue of the
presence of instabilities of the kind we are interested in. The
classical potential on the space of these configurations is
\EQ{S^{(0)}=\beta R \pi^2 {1\over g^2_{YM}}\;\Tr\varphi_a^2.}

Hence, as indicated above, for the effective potential calculation we 
shift $\phi_a\to\varphi_a+\phi_a$ where we assume that 
$\{\varphi_a\}$ and the Polyakov loop $\alpha$ are simultaneously
diagonal. The fluctuations $\{\phi_a\}$ which will be
integrated out are fluctuations about the background, comprising of 
all the non-constant
modes on $\sph{3}\times \sph{1}$ and 
off-diagonal components of the homogeneous modes.

We denote indices along $\sph{3}$ as $i=1,2,3$ (not to be confused
with the gauge index) and that along
$\sph{1}$ as $0$. 
Unlike \cite{Aharony:2003sx} who consider the case with vanishing
background fields, 
we find it easier to work in a covariant gauge. In
particular, we choose a conventional ``$R_\xi$ gauge'' of a 
spontaneously broken gauge
theory. To this end, we add to the action the gauge fixing term
\EQ{
S_\text{gf}={1\over g^2_{YM}}\frac1{2\xi}\int
d^4x\sqrt{g}\,\text{Tr}\Big(\nabla_iA^i+\tilde 
D_0A_0-i\xi\varphi_a\phi_a\Big)^2,\
\label{gfa}
}
which implements a covariant gauge condition. Here and in what follows
$\tilde D_0=\partial_0+i\alpha$ includes $\alpha$, the zero mode part
of $A_0$ only. Also, we leave adjoint action by the background fields
$\varphi$ and $\alpha$ as implicit,   
{\it i.e.}~$\varphi\phi\equiv[\varphi,\phi]$,
$\varphi^2\phi\equiv[\varphi,[\varphi,\phi]]$, {\it etc.}

We will now expand the gauge fixed action including the ghosts to
quadratic order in fluctuations about the background fields $\varphi_a$
and the Polyakov loop $\alpha$. The one loop determinants obtained by
integrating out these fluctuations yield the radiative
corrections to the effective potential.
 In order to proceed,
it is particularly convenient to choose the Feynman
gauge $\xi=1$. In this case the action for the bosonic quadratic
fluctuations, including ghosts, takes a simple form:

\SP{
S_{\rm b}=&\frac1{g^2_{YM}}\int
d^4x\sqrt{g}\,\text{Tr}\,\Big[\tfrac12
A_0(-\tilde D_0^2-\Delta^{(s)}+\varphi_a^2)A_0
+\tfrac12A_i(-\tilde D_0^2-\Delta^{(v)}+\varphi_a^2)A^i\\ &+
\tfrac12\phi_a(-\tilde D_0^2-\Delta^{(s)}+R^{-2}+\varphi_b^2)\phi_a
+\bar c(-\tilde D_0^2-\Delta^{(s)}+\varphi_a^2)c\Big]\ .
\label{asd}
}
Here, $\Delta^{(s)}$ and $\Delta^{(v)}$ are the Laplacians
on $S^3$ for
scalar and gauge fields, respectively. The scalar Laplacian is simply
$\nabla_i\nabla^i$ acting on scalar functions whilst the vector
Laplacian is defined via the quadratic terms in the action:
\begin{eqnarray}
&&\int d^4x\sqrt{g}\,\;\text{Tr}\,\tfrac14F_{\mu\nu}F^{\mu\nu}=
\label{vgg}\\\nonumber
&&\int d^4x\sqrt{g}\,\text{Tr}\big(-\tfrac12
A_\mu(\tilde D_0^2+\Delta^{(v)})A^\mu-
\tfrac12(\tilde D_0A_0+\nabla_i A^i)^2+{\cal O}(A^3)\big).
\end{eqnarray}
The second term on the right hand side is canceled by terms in the
gauge fixing action 
\eqref{gfa}. From \eqref{vgg}, it follows that
\EQ{
\Delta^{(v)}A^i=\nabla_j\nabla^jA^i-R^i{}_jA^j\ ,
}
where $R_{ij}$ is the Ricci tensor of $S^3$. For the temporal
component of the gauge field, the Ricci tensor does not contribute and
the vector 
Laplacian is equivalent to the scalar Laplacian:
\EQ{
\Delta^{(v)}A_0=\Delta^{(s)}A_0\ .
}
The eigenvectors of the scalar Laplacian are spherical harmonics $Y_\ell$
labeled by angular momentum $\ell\in{\mathbb Z}\geq0$ with 
\EQ{
\Delta^{(s)}Y_\ell=- \ell(\ell+2)\,R^{-2}\,Y_\ell, 
}
and degeneracy
$(\ell+1)^2$. The eigenvectors of the vector Laplacian can be split
into two sets. Firstly, those in the image of $\nabla^i$, {\it
  i.e.\/}~$\nabla^iY_\ell$ with $\ell>0$, which satisfy
\EQ{
\Delta^{(v)}\nabla^iY_\ell=- \ell(\ell+2)\,R^{-2}\,\nabla^iY_\ell\ ,
\label{v1}}
and secondly those in kernel, $\nabla_iV^i_\ell=0$, also labeled by the
angular momentum $\ell>0$, with
\EQ{
\Delta^{(v)}V^i_\ell=- (\ell+1)^2\,R^{-2}\,V^i_\ell, 
\label{v2}}
and degeneracy $2\ell(\ell+2)$.

We are now in position to integrate out all the bosonic fluctuations.
 For the vector modes,  we must first split the $A^i$ into the image and the
 kernel of $\nabla^i$. To this end we write $A^i=B^i+C^i$ with
 $\nabla_iB^i=0$ and  
$C^i=\nabla^i f$. Integrating these out along with $A_0$, the ghosts
 and the six scalar fluctuations yields the following contribution to
 the one loop effective potential from bosonic radiative corrections:
\SP{& S^{(1)}_{\rm b}=
\tfrac12\log\det_{\ell>0}(-\tilde D_0^2
-\Delta^{(v)}+\varphi_a^2)+\tfrac12\log\det_{\ell>0}(-\tilde D_0^2
-\Delta^{(s)}+\varphi^2)+\\\\
&+(\tfrac12-1)\log\det_{\ell\geq0}(-\tilde D_0^2
-\Delta^{(s)}+\varphi_a^2)+6\times\tfrac12\log\det_{\ell\geq0}(-\tilde D_0^2
-\Delta^{(s)}+R^{-2}+\varphi_a^2)\ .
\label{s1}
}
We now describe the various contributions in some detail. 
The first two terms arise from integrating out the vector fluctuations
$B_i$ and $C_i$ respectively where the subscripts
remind us to exclude the $\ell=0$ mode of $\Delta^{(s)}$ as discussed
above equations \eqref{v1} and \eqref{v2}. The third term in the
effective potential results from the integrals over the ghost and $A_0$
fields, each of which contribute the same factor with weight $-1$ and
$\tfrac12$ respectively. Finally we have the fluctuation determinants
of the six scalar fields of the $\N=4$ theory.  It is vital to realize that
the determinants for the ghosts, the scalars and $A_0$  all
include the $\ell=0$ modes as indicated.
Note that the fluctuation determinants for the $C_i$, the ghosts and
$A_0$ combine neatly due to a complete cancellation between non-zero
momentum modes yielding a simplified expression for the bosonic
one-loop correction:

\SP{S^{(1)}_{\rm b}=
&\tfrac12\log\det_{\ell>0}(-\tilde D_0^2
-\Delta^{(v)}+\varphi_a^2)-\tfrac12\log\det_{\ell=0}(-\tilde
D_0^2+\varphi_a^2)\\\\
&+6\times\tfrac12\log\det_{\ell\geq0}(-\tilde D_0^2 
-\Delta^{(s)}+R^{-2}+\varphi_a^2)\ .
\label{s2}
}
\\
We can make these formal expressions explicit by rewriting each of the
terms above as a trace over the thermal frequencies or
Matsubara modes and the discrete wave numbers on $\sph{3}$. 
The result of the Matsubara trace of a typical contribution in the
effective action takes the form (up to field independent additive constants)
\EQ{
\text{Tr}\log(-\tilde D_0^2+\varepsilon^2)= \text{Tr}\Big(\beta\varepsilon-2
\sum_{n=1}^\infty\frac1ne^{-\beta\varepsilon}\cos(n\beta\alpha)\Big)\ .
}
\\
Here $\varepsilon$ is the energy of the fluctuation in question 
with $x^0$ interpreted as
imaginary time and we have used the fact that the operator $\tilde
D_0$ has eigenvalues $(2\pi n/\beta+\alpha)i$; $n\in{\mathbb Z}$. 
The remaining trace on the right hand side is to be taken over
the modes on $\sph{3}$ and the gauge group. 
The gauge trace can be made explicit by using the fact that 
$\alpha$ and $\varphi$ were chosen to be simultaneously diagonal
\EQ{
\alpha=\beta^{-1}\text{diag}\;(\theta_1,\theta_2,\ldots\theta_N) ,\qquad
\varphi=\text{diag}\;(\varphi_1,\varphi_2,\ldots\varphi_N)\ ,
}
which in turn yields in general
\EQ{
\text{Tr}\,f(\varphi,\alpha)=\sum_{ij=1}^Nf(\varphi_{ij},\alpha_{ij});\quad
\varphi_{ij}=\varphi_i-\varphi_j\;,\quad 
\alpha_{ij}=\;\beta^{-1}(\theta_i-\theta_j).
}
For $SU(N)$ gauge group the $\theta_i$ and $\varphi_i$ must satisfy
\EQ{\sum_{i=1}^N\varphi_i=0\;,\qquad\sum_{i=1}^N\theta_i=0\;{\rm mod}
\;2\pi.\label{periodicity}}
These conditions are a consequence of the Hermiticity of the scalar
and gauge fields. The $\theta_i$ are each thermal Wilson lines for the
Cartan components of the gauge field. Hence they can
be shifted by an integer multiple of $2\pi$ by performing a
topologically non-trivial gauge transformation which is single-valued
in $SU(N)/{\mathbb Z}_N$. This must however be an invariance of the
theory since there are no fields charged under the center ${\mathbb
  Z}_N$ of $SU(N)$. This results in the $\theta_i$ being defined only
up to an integer multiple of $2\pi$.

Including the explicit sums over the angular momenta on $\sph{3}$ with
the appropriate degeneracies, we 
 can now write down the complete one-loop effective action resulting
 from integrating out all the bosonic fluctuations, 

\SP{S_{\rm b}^{(1)}\;&=\;\sum_{ij=1}^N\Big[
-{1\over 2}\beta\sqrt{\varphi_{aij}^2}+
\sum_{n=1}^\infty\frac1ne^{-n\beta\sqrt{\varphi_{aij}^2}}\cos(n\theta_{ij})
\\
+&\sum_{\ell=1}2\ell(\ell+2)\Big(
\frac\beta2
\sqrt{ (\ell+1)^2\,R^{-2}\,+\varphi_{aij}^2}-\sum_{n=1}^\infty\frac1ne^{-n\beta
\sqrt{ (\ell+1)^2\,R^{-2}\,+\varphi_{aij}^2}}\cos(n\theta_{ij})\Big)
\\
+&\;6\sum_{\ell=0}(\ell+1)^2\Big(
\frac\beta2\sqrt{ (\ell+1)^2\,R^{-2}\,+\varphi_{aij}^2}
-\sum_{n=1}^\infty\frac1ne^{-n\beta
\sqrt{ (\ell+1)^2\,R^{-2}\,+\varphi_{aij}^2}}\cos(n\theta_{ij})\Big)
\Big]. \label{sbosonic}}
Finally, to complete the calculation of the effective action we have to include
integrals over the four Weyl fermions. 
The effect of the background fields is very simple on these modes, it
simply induces a mass squared $\varphi_a^2$ for the fermions via their
Yukawa couplings.  
The fermions are eigenfunctions of the spinor Laplacian 
on $\sph{3}$ which are also
labeled by the angular momentum $\ell>0$ with eigenvalue 
$- (\ell+\tfrac12)^2$ and degeneracy $2\ell(\ell+1)$. 
The key difference between fermions and bosons at finite temperature
is of course that the former obey anti-periodic boundary conditions
around the thermal circle. Thus, when acting on fermions the
$\tilde D_0$ operator has eigenvalues $(2\pi
(n+{1/2})/\beta+\alpha)i$; $n\in {\mathbb Z}$.

Hence, with the anti-periodic boundary conditions around $\sph{1}$ the
fermionic 
contribution to the effective action is 
\SP{
S_\text{f}^{(1)}=
4\sum_{ij=1}^N\sum_{\ell=1}^\infty & 2\ell(\ell+1)\Big(
-\frac\beta2\sqrt{ (\ell+\tfrac12)^2\,R^{-2}\,+\varphi_{aij}^2}
\\ &-\sum_{n=1}^\infty\frac{(-1)^{n+1}}ne^{-n\beta
\sqrt{ (\ell+\tfrac12)^2\,R^{-2}\,+\varphi_{aij}^2}}\cos(n\theta_{ij})\Big).
\label{sfermionic}}

The complete one loop effective potential for the eigenvalues of the
Polyakov 
loop variable $\alpha$ and those of the constant modes $\varphi_a$ is
given by 

\EQ{S_{\rm eff}[\varphi_{ai}, \theta_i]\;=\; S^{(0)}+S^{(1)}_{\rm b} +
  S^{(1)}_{\rm f}, \label{full}}
\\
where the tree level potential for the eigenvalues is
\EQ{S^{(0)}=\beta R\pi^2{1\over g^2_{YM}}\; \sum_{i=1}^N \varphi_{ai}^2,}
while $S^{(1)}_{\rm b}$ and $S^{(1)}_{\rm f}$ are determined in
\eqref{sbosonic} and \eqref{sfermionic}. Note the power of
  $1/g^2_{YM}$ in front of the tree level term indicating that our one
  loop contributions are parametrically down by a power of the weak
  gauge coupling.  
\begin{table}
\begin{center}\begin{tabular}{ccccc}\hline\hline
field & angular mom. & energy & degeneracy & weight \\
\hline
$B_i $ & $ \ell>0 $ & $ \sqrt{(\ell+1)^2\,R^{-2}\,+\varphi^2} $ & $
2\ell(\ell+2) $ & $ \tfrac12$ \\ 
$C_i $ & $ \ell>0 $ & $ \sqrt{\ell(\ell+2)\,R^{-2}\,+\varphi^2} $ & $
(\ell+1)^2 
$ & $ \tfrac12$ \\ 
$\bar c,c $ & $ \ell\geq0 $ & $ \sqrt{\ell(\ell+2)\,R^{-2}\,+\varphi^2} $ & $
(\ell+1)^2 $ & $ -1$ \\ 
$A_0$ & $ \ell\geq0 $ & $ \sqrt{\ell(\ell+2)\,R^{-2}\,+\varphi^2}$ & $
(\ell+1)^2 $ & $ 
\tfrac12$ \\
$\phi_a $ & $ \ell\geq0 $ & $ \sqrt{(\ell+1)^2\,R^{-2}\,+\varphi^2} $
& $ (\ell+1)^2 
$ & $ \tfrac12 $ \\
$\psi^A_\alpha $ & $ \ell>0 $ & $
\sqrt{(\ell+\tfrac12)^2\,R^{-2}\,+\varphi^2} $ & $ 2\ell(\ell+1) $ & $
-\tfrac12$ \\
\hline\hline
\end{tabular}\end{center}
\caption{\small The fields, their angular momenta, energy, degeneracy
  and their weight in the effective action.}
\end{table}

Table 1 summarizes the properties of all the 
modes including their range of angular momenta, energies,
degeneracies and the weight of the corresponding ``$\log\det$'' terms in
the effective potential. 
\\\\
\\\
\underline{\it Discussion}
\\\\
We now make the following observations about our result: 
\begin{itemize}
\item{Firstly,
when the scalar fields $\varphi_a$ vanish we reproduce the
results of \cite{Aharony:2003sx}. To see how this works, we note that
the first term in the one-loop effective potential in the absence of
the background scalars reduces to 
\EQ{
\log\prod_{i<j}\sin^2\Big(\frac{\theta_{ij}}2\Big)\ ,
}
which is precisely the Jacobian that converts the integration measure over the
Hermitian matrix $\alpha$ into the appropriate measure for the Unitary
matrix $U=e^{i\beta\alpha}$,
\EQ{
\int[d\alpha]\,\det(-\tilde D_0)=\int dU=
\int\prod_{i=1}^Nd\theta_i\,\prod_{i<j}\sin^2\Big(\frac{\theta_{ij}}2\Big)\ .
}
Here we have obtained this result following a different route from
that adopted by \cite{Aharony:2003sx}.
So when the scalars vanish it is easy to
see that we reproduce the effective action written down in 
\cite{Aharony:2003sx}.}

\item{Our effective potential was obtained by integrating out not only
the non-zero momentum modes on $\sph{3}\times\sph{1}$, but also the
off-diagonal fluctuations of $\alpha$ and of the zero modes ($n=\ell=0$) of
$\phi_a$. (In principle one also integrates out fluctuations of the
diagonal pieces but these give no nontrivial contributions, due to the
adjoint action of the background fields on them). Note that the 
off-diagonal zero modes have masses
$\sqrt{R^{-2}+\sum_{a}(\varphi_{ai}-\varphi_{aj})^2}$ which are 
generically larger than the tree level masses for the diagonal
components of the constant modes.
Accounting for these constant modes is precisely what gives
rise to the  
appropriate Jacobian factor for unitary matrices discussed above and
the repulsive  
Vandermonde interaction between the eigenvalues $\varphi_{ai}$ of the
scalars. The presence of the logarithmic repulsive force between the
$N$ eigenvalues $\varphi_{ai}$ is easily inferred from the fluctuation
determinant for the  $\ell=n=0$ modes in \eqref{s2}. }

\item{The potential is only a function of the variable 
\EQ{\varphi_{aij}^2=\sum_{a=1}^6(\varphi_{ai}-\varphi_{aj})^2.}
A non-zero expectation value for this will single out a particular
direction in the space spanned by the six scalars and hence will
spontaneously break the $SO(6)$ R-symmetry to $SO(5)$ in the large $N$
theory. Note that the notion
of symmetry breaking on a compact space is unusual and we clarify this
carefully in subsection 3.3.2.} 

\item{The thermal interpretation of the various terms in the effective
potential is also fairly clear. All the finite temperature
contributions are accompanied by thermal Boltzmann suppression
factors. In the zero temperature limit $\beta=\infty$, these vanish
leaving behind the terms which are manifestly linear in
$\beta$. The latter are the zero temperature or ``Casimir energy''
contributions on $\sph{3}$ in the presence of scalar expectation values. We
will discuss these zero temperature contributions in more detail in the following
section. For now, we make one additional observation that the
term proportional to $\sqrt{\varphi_{aij}^2}$ actually cancels
against an identical piece emerging from the infinite sums over $\ell$.}

\item{The bosonic and fermionic determinants come with opposite
signs as they must. It is, however, important to realize that they do
  not cancel against each other due both to finite temperature effects
  and the fact that the theory is formulated on a spatial
  three-sphere. A complete cancellation between bosonic and fermionic
  fluctuations only occurs in flat space and at zero temperature. We
  will come back to this point in the following subsection.}

\item{The effective potential that we have obtained can be interpreted as a
1PI-effective action for the eigenvalues of $\alpha$ and the
$\varphi_a$ and yields the one loop partition function about a given
configuration of eigenvalues. At local minima of the action, we may
interpret the partition function as a thermodynamic partition function
whose logarithm is the thermodynamic free energy. For generic points
in the configuration space of field eigenvalues, the system will not
be in a stable or static configuration. Nevertheless, we may  
formally define a free energy as
\EQ{
F(\varphi,\theta)=S_\text{eff}/\beta.}

We are interested both in extrema of this
effective potential and the existence of and emergence of any new
unstable directions  
as a function of temperature.}

\item{Finally, saddle points computed using the effective potential
   will be true saddle points in the full configurations space, even
   though we are setting to zero the background values of the
   off-diagonal constant modes. The reason for this is that, in our
   chosen background of diagonal fields, the off-diagonal fluctuations
   only appear quadratically and there are no terms linear in these
   fluctuations. This means that there are extrema of the full
   effective action where these can be consistently set to zero and
   these are the saddle points that we will find. However, there may
   well be other extrema where the fields $\varphi$ and $\alpha$ are not
   simultaneously diagonal. }
\end{itemize}

\subsection{The zero temperature limit}

In this section we discuss the detailed form of the one loop effective
potential at zero temperature. This is not relevant for the high
temperature analysis which is our main focus. However, the limit of
zero temperature will provide checks of our calculation and also
important intuition about the behaviour of the one-loop corrections.

The terms that survive in the zero temperature limit, 
$\beta=\infty$, are basically Casimir contributions in the presence of
background expectation values, and are identified as the terms without
any Boltzmann 
suppression factors in the effective potential. 
The total Casimir free energy is therefore
\EQ{
F_0\equiv\sum_{ij=1}^NC(\varphi_{aij}^2)\ ,
}
where
\SP{
C(\varphi^2)&=\sum_{\ell=1}^\infty\ell(\ell+2)\sqrt{
(\ell+1)^2\,R^{-2}\,+\varphi^2}+
3\sum_{\ell=0}^\infty(\ell+1)^2\sqrt{(\ell+1)^2\,R^{-2}\,+\varphi^2}\\&
-4\sum_{\ell=1}^\infty\ell(\ell+1)\sqrt{(\ell+\tfrac12)^2\,R^{-2}\,+\varphi^2}
-{1\over 2}\varphi\ .
\label{gtt}
}
Note the complete absence of any dependence on the Polyakov loop
variable $\alpha$.

Furthermore, as we remarked earlier,  the bosonic and fermionic
determinants do not cancel against each other even at zero
temperature though the system is supersymmetric. This is purely a
consequence of formulating the theory on a curved spatial manifold,
namely an $\sph{3}$. Hence there is a nontrivial quantum correction to
scalar potential of the ${\cal N}=4$ theory on $\sph{3}\times{\mathbb
  R}$ even though all supersymmetries are unbroken. The cancellation
is recovered in the flat space limit.

Evidently the infinite sums, at least individually, require
regularization. This can be done by introducing a generalized
$\zeta$-function, or simple Epstein series,
\EQ{
{\cal E}(z,\theta;\varphi R)=
\sum_{n=-\infty}^\infty\big[(n+\theta/2\pi)^2+(\varphi R)^2\big]^{-z}\ .
} 
In terms of this, the Casimir energy can be neatly expressed as
 \SP{C(\varphi^2)=
2\,R^{-1}\,\Big[{\cal E}(-\tfrac32,0;\varphi R)&-{\cal
       E}(-\tfrac32,\pi;\varphi R)\\
&-((\varphi R)^2+{\tfrac14})\left({\cal E}(-\tfrac12,0;\varphi R)-{\cal
       E}(-\tfrac12,\pi;\varphi R) \right)
\Big].\label{ce}}
The bosonic and fermionic contributions mirror each other with
opposing signs, the former yielding Epstein series with $\theta=0$
while the latter have $\theta=\pi$. 

It is easy to extract the zero point energy at $\varphi=0$ formally in
terms of the Riemann zeta function and the generalized Riemann zeta function 
\SP{F_0[0]=&{2 \over R} \Big[2
    \zeta(-3)-\zeta(-3,\tfrac12)-\zeta(-3,-\tfrac12)
-\tfrac14\Big(2 \zeta(-1)-\zeta(-1,\tfrac12)-\zeta(-1,\tfrac12)\Big)
\Big]\\
=&{3\over 16R }~.
}
This can be identified as
\EQ{
\frac3{16}=
6\times\frac1{240}+\frac{11}{120}+4\times\frac{17}{960},
}
the well-known sum of the zero-point energies of
scalars, vectors and fermions on a three-sphere appropriate to the
${\cal N}=4$ theory.

More interesting is the behaviour for generic nonzero $\varphi_{aij}$
wherein we need the analytic
continuation of the Epstein series which is known to be (see for example
\cite{Elizalde:2003ke} or \cite{Cavalcanti:2006yd})
\EQ{
{\cal E}(z,\theta;\varphi R)
=\frac{\sqrt\pi \,(\varphi R)^{1-2z}}{\Gamma(z)}
\Big[\,\Gamma(z-1/2)+4\sum_{n=1}^\infty\cos(n\theta)
\,\frac{K_{1/2-z}(2n\pi\varphi R)}{(\pi n\varphi R)^{1/2-z}}\Big]\ .
\label{gt}
}
The singularities corresponding to the individually divergent sums we
encountered earlier appear as the poles of the gamma
function $\Gamma(z-1/2)$ for $z=-\tfrac12$ and $-\tfrac32$. Notice
that these divergent poles correspond to putative mass and coupling constant
renormalizations respectively since they are coefficients
of $\varphi^{1-2 z}$. However, the magic of the $\N=4$ theory (whose
supersymmetry is recovered in the limit $\beta\to\infty$) ensures that
these poles cancel against each other in \eqref{ce} to leave a 
finite result. Hence no subtraction or renormalization is necessary,
reflecting the UV finiteness of the theory.

The resulting expression for the zero temperature effective potential
is then 
\EQ{F_0[\varphi_{aij}]=\sum_{ij=1}^N C(\varphi_{aij}^2),}
with 
\SP{
&C(\varphi^2)=\\&
\frac{12(\varphi R)^2}{\pi^2 R}\;\sum_{n=1}^\infty
\frac{K_2(2(2n-1)\pi\,\varphi R)
}{(2n-1)^2}+\frac{2\varphi R\,(4 (\varphi R)^2+1)}
{\pi R}\sum_{n=1}^\infty\frac{K_1(2(2n-1)\pi \varphi R)}{2n-1}~,
}
which manifestly vanishes exponentially for large $\varphi$ using 
standard asymptotic properties of the Bessel functions. The reason we
expect these one loop effects to vanish at large $\varphi R$ is
essentially a decoupling argument. The masses of the modes being
integrated out increase with increasing $\varphi$ and hence their 
contributions are suppressed at large $\varphi$.

For small $\varphi$, the expansion of $C(\varphi)$ is poorly behaved; 
nevertheless its asymptotic limit $\varphi R\to0$ can be extracted:
\EQ{
C(0)=\frac{6}{\pi^4 R}\sum_{n=1}^\infty\frac1{(2n-1)^4}
+\frac{1}
{\pi^2 R}\sum_{n=1}^\infty\frac1{(2n-1)^2}=\frac3{16R}\, ,
}
coinciding with the $\varphi=0$ result above.

\begin{figure}[ht] 
\centerline{\includegraphics[width=3in]{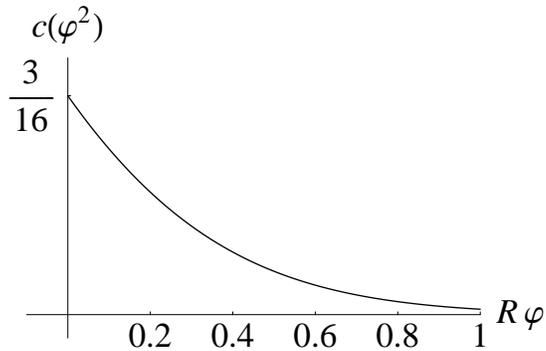}}
\caption{\footnotesize The one-loop correction at zero
  temperature on $S^3$.}  
\label{casimir}
\end{figure}

A numerical approximation of the 
zero temperature effective potential is plotted in
Figure~\ref{casimir}. The full potential (along the simultaneously
diagonal scalar directions) is obtained by adding to it, the tree level
conformal coupling to the curvature.

\subsection{High temperature effective potential}

We now turn to the high temperature behaviour of the effective
potential. As is well-known, the high temperature behaviour of
gauge theories is subtle even at weak gauge coupling requiring a
detailed understanding of the relevant length scales in
question.

In the high temperature limit for the $\N=4$ theory, 
 the temperature is much larger than
the inverse radius of the $\sph{3}$
\EQ{T \gg {1\over R}~.}
Naively this is like a flat space limit. However the situation is
more subtle since the theory has an infrared cutoff corresponding to
the finite size of the three-sphere and what we really need to know is
how the combination $T R$ scales with the weak ''t Hooft coupling
$\lambda=g^2_{YM} N$. We will come back to this issue later. For now,
we note that a weakly coupled $SU(N)$ Yang-Mills plasma has a
hierarchy of length 
scales (see e.g \cite{kapusta, lebellac}), chief among these being the
temperature $T$, followed 
by the perturbative electric screening scale or Debye length $
(\sqrt\lambda T)^{-1}$ and the non-perturbative magnetic screening
length $(\lambda T)^{-1}$. Each length scale is described by an
appropriate effective theory obtained by integrating out higher
momentum modes. 


%

\begin{figure}[ht] 
\centerline{\includegraphics[width=2.5in]{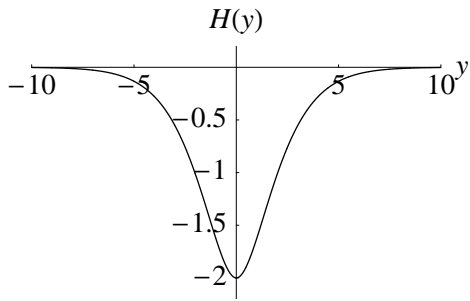}}
\caption{\footnotesize The function ${\cal H}(y)$.} \label{bessel1}
\end{figure}

Let us first take the naive high temperature limit $(TR\gg1)$ of our one loop 
effective potential \eqref{full}. In this
limit, all the mode sums over $\ell$ 
can be approximated by momentum integrals over a continuous
variable  $x=n\ell/(TR)$. In addition, the zero temperature
contributions to the one loop potential are sub-leading and the
high temperature effective action takes the form,
\\
\SP{&S_{\rm eff}[\phi,\theta]= \\
&2 \beta R^3\pi^2 \left[
{R^{-2}\over  2 g^2_{YM}}\;\sum_{i=1}^N\varphi_{ai}^2 \;\;+\;\;4 {T^4\over
  \pi^2}\;\sum_{n=1}^\infty \sum_{ij=1}^N (1-(-1)^n) {1\over n^4}
{\cal H}\left(n{|\varphi_{aij}|\over
    T}\right)\cos(n\theta_{ij})\right].\label{highT}} 
where 
\EQ{|\varphi_{aij}|\equiv\left({\sum_{a=1}^6}\varphi_{aij}^2\right)^{1\over
  2}.}
We have expressed our result in terms of the function
\EQ{
{\cal H}(y)=-\int_0^\infty dx\,x^2e^{-\sqrt{x^2+y^2}}
=-\frac{y}2\big(y K_0(y)+2K_1(y)+
yK_2(y)\big)\ ,
}
which is plotted in \eqref{bessel1}. 
Notice that it is a negative definite, even, monotonically increasing
function of $|y|$. For small $y$ the function ${\cal H}(y)$ approaches
a constant as 
 \EQ{
\lim_{y\to0}{\cal H}(y)=-2+\frac{y^2}2+{\cal O}(y^3)\ ,
\label{asz}
}
while as $y\to\infty$, ${\cal H}(y)$ approaches zero exponentially as  
\EQ{
\lim_{y\to\infty}{\cal H}(y)=-\sqrt{\frac\pi2}y^{3/2}e^{-y}\ .
\label{asi}
}

The limit we have taken yields an $R$-independent expression for the 
one loop term in \eqref{highT}. Consequently 
we obtain the flat space result
for a combined one loop potential for the scalars and the Polyakov
loop in thermal $\N=4$ theory. \footnote{Similar expressions were
  obtained by \cite{Yamada:2006rx} but for the case where the Polyakov
  loop variables are set to zero, corresponding to the deconfined
  phase, along with the inclusion of a finite chemical potential.} 
The only trace of the curved
background is in the tree level mass term originating from coupling to
the background curvature. Strictly speaking we should include
an infrared cutoff in the momentum integrals reflecting the
compactness of the three-sphere background, although it will not alter
our result in the regime of interest. The regime of validity of the
above potential will be discussed in more detail below.

We will now explore some general properties of
the high temperature potential obtained above. The one loop
correction is only a  
function of the differences $\varphi_{aij}=\varphi_{ai}-\varphi_{aj}$,
and the eigenvalues $\varphi_{ai}$ each experience a quadratic
classical potential due to the mass. Therefore for $U(N)$ gauge group
we can consistently set
to zero the expectation value of the scalar belonging to
the diagonal $U(1)$ multiplet of the $U(N)$ 
\EQ{\sum_{i=1}^N{\varphi_{ai}=0}~.}
Alternatively, this is automatically true if we work with $SU(N)$
gauge group from the start. Using this, and defining the dimensionless field
\EQ{\tilde{\varphi}_{aij}\equiv {{\varphi_{ai}-\varphi_{aj}}\over T},}
the effective action assumes the form
\\
\SP{&S_{\rm eff}=\\&
2(TR)^3\pi^2 \sum_{ij=1}^N\left[
{R^{-2}\over \lambda T^2}\;\tilde{\varphi}_{aij}^2
\;\;+\;{1\over \pi^2}\;\sum_{n=1}^\infty 
(1-(-1)^n) {4\over n^4}
{\cal H}(n{|\tilde{\varphi}_{aij}}|)\cos(n\theta_{ij})\right],\label{fulleff}}
\\
where the 't Hooft coupling $\lambda=g^2_{YM} N$ makes a natural
appearance. As before we have used the shorthand  $|\tilde
\varphi_{aij}|$ for $(\sum_a\varphi_{aij}^2)^{1\over 2}$.
Note the emergence of the dimensionless ratio involving the Debye
mass scale
\EQ{{R\over {(\sqrt\lambda T)^{-1}}}~,}
which governs the size of the quantum correction relative to the
classical term. The classical piece will dominate in the high
temperature limit provided
\EQ{1\ll TR < {1\over\sqrt\lambda}~, \label{range}}
and perturbation theory in $\lambda$ will be valid in this
regime. In other words, for perturbation theory in $\lambda$ to make
sense, the size of the three-sphere must not exceed
the Debye screening length scale. But since we are at arbitrarily weak
coupling, we can always choose high enough temperature while
satisfying this requirement so that the discrete mode sums
can be effectively replaced with integrals.

Note also that since one loop contributions are planar, the quantum
corrections we have computed automatically scale as $N^2$ in the 't
Hooft large
$N$ limit, due to the double summation over $i$ and $j$ indices.
 
\subsubsection{A global minimum and big AdS black hole}
The function ${\cal H}(y)$ is negative for all values of
$y$. Hence the absolute minimum of $S_{\rm eff}$ 
is obtained when  all $\varphi_{ai}$ vanish  \footnote{ Actually, there is repulsive Vandermonde type interaction which leads to an eigenvalue spread of $O(\lambda)$ about $\varphi=0$. This is quantum effect and subleading compared to the effects that are of interest to us.} and all the $\theta_{i}$ are
equal. Using  \eqref{periodicity}, $\sum_i \theta_i=0 ~{\rm mod} ~ 2
\pi$, we find a  non-zero expectation value for the Polyakov 
loop:
\EQ{\langle U \rangle \;\equiv\; { 1 \over N} \Tr \;e^{i\beta
    \alpha}\;=\;{1\over N} \sum_{j=1}^N e^{i\theta _j}\;=\;  
e^{{2\pi i\over N} k},\qquad k=1,2,\ldots N.}
Actually, in finite volume, one must define this expectation value
carefully as discussed in \cite{Aharony:2003sx} in order to avoid a vanishing
result due to averaging over the ${\mathbb Z}_N$ vacua. To this end
one first introduces a
bias which breaks the ${\mathbb Z}_N$ symmetry, then takes a large $N$ limit
and subsequently removes the bias to obtain a non-zero value for
$\langle U\rangle$. Alternatively, one may simply compute 
$\langle |U|^2\rangle$.
A non-zero Polyakov loop implies that the theory is in a high
temperature deconfined phase without any VEVs for scalars. Indeed,
this saddle point of the effective potential 
coincides with the high temperature deconfined phase of the free
$\N=4$ theory identified in \cite{Aharony:2003sx}. 

Since we have argued that this is a global minimum of the action, we
know that the saddle point is stable, at least within the range of
temperatures in which the one loop approximation is valid. Importantly
for us, this deconfined phase is the continuation to weak coupling of
the large AdS Schwarzschild black hole which is the gravity dual of the
strongly coupled deconfined plasma phase. At high temperatures, this
minimum will dominate the canonical ensemble. 

Let us finally check that this deconfined vacuum is indeed stable to
fluctuations in the scalar directions. Using the behaviour of ${\cal H} (y)$
near the origin, we find
\EQ{S_{\rm eff} \simeq \; \pi^2 R^3 \beta
\sum_{ij=1}^N\left[\;{R^{-2}\over\lambda}
\varphi_{aij}^2\;+ \;T^2 \;\varphi_{aij}^2\;\right], \qquad{\rm as}\qquad
{\varphi_{aij}\over T }\;\rightarrow\;0 ,}
which is clearly stable for all temperatures in the range above and
formally for all $T$. What we have found here is simply the
finite temperature mass renormalization or the thermal mass of the
scalars in the deconfined phase. As anticipated earlier the one loop thermal
mass in the deconfined phase is given by
\EQ{m_{\rm th}=\sqrt{\lambda} T~.}
This stability of the deconfined plasma phase at weak coupling is also
in line with the 
absolute stability of the big AdS Schwarzschild black hole at strong coupling.

\subsubsection{Unstable directions charged under R-symmetry group}
So far we have restricted ourselves to a regime of temperatures
\eqref{range} wherein perturbation theory in $\lambda$ is valid and
where the classical term dominates the first quantum correction. In
this temperature range, for arbitrary
$\{\theta_i\}$ the classical mass ensures that small
$\varphi$-fluctuations are locally stable.
\\\\
\underline{\it Perturbation theory at high temperature}

Importantly, the one loop result actually has a much wider range of
validity.  It is in fact also valid at temperatures where it 
competes with the tree level term, {\em i.e.}
\EQ{TR\sim{1\over \sqrt\lambda},}
while higher order perturbative contributions remain consistently
parametrically small. The reason for this is fairly well-known in flat
space: At
length scales larger than the Debye screening  length and smaller than
the magnetic screening length, perturbation theory remains a good description
with one key difference -- the expansion parameter is $\sqrt\lambda$
instead of $\lambda$. Of course, the situation is simpler on 
$\sph{3}$ with $R\sim(\sqrt\lambda T)^{-1}$ providing an infrared
cutoff.

In the present context we may understand this as follows.
In the high
temperature limit, replacing discrete mode sums with integrals, 
the $n$ loop contribution to the effective
potential in naive perturbation theory will be of the form
\EQ{{\tfrac1{{\rm Vol}(\sph{3}\times
      \sph{1})}}\;S^{(n)}[\varphi]\;\sim\;\lambda^{n-1}\;T^4 
    \;F_n\left({\varphi\over T}\right)} 
simply by dimensional analysis and power counting. For $n>1$ these
clearly appear to be sub-leading to the one loop term at any
temperature, including when the one loop effect competes with the
classical piece. However, a potential subtlety arises as
${\varphi/T}\rightarrow 0$. In flat space, at two loops and at higher
orders, in this limit one encounters infrared divergences which
necessitate a resummation of the thermal mass of the excitations. In
our case, while we should consider a similar resummation, an infrared
cutoff is already provided by the size of the three-sphere. When 
$R\sim (\sqrt\lambda T)^{-1}$ this IR cutoff is the 
Debye scale and perturbation theory continues to be valid albeit in
the parameter $\sqrt\lambda$ as opposed to $\lambda$. Operationally
this is automatically achieved since the energies of the lowest $\ell=0$
bosonic and fermionic modes are $\sqrt{R^{-2}+\varphi^2}$ and 
$\sqrt{\tfrac14 R^{-2}+\varphi^2}$ (see Table 1). Upon inclusion of this
  IR cutoff in our momentum integrals, 
the two loop contribution as $\varphi/T\rightarrow 0$ scales with an
extra power of $\sqrt\lambda$ relative to the one loop term and is
suppressed at weak 't Hooft coupling.

In summary then, within the regime of validity of perturbation theory
we are free to consider temperatures where the tree level and one loop
terms become comparable. 
\\
\\
\\
\underline{\it Unstable directions and a critical temperature}
\\\\
In the temperature range $R^{-1}\ll T< {1\over\sqrt\lambda}R^{-1}$,
for any configuration of $\theta_i$'s, fluctuations in the
$\varphi$ direction are stable since the classical mass term dominates. 
Within this regime,
the global minimum of the potential
occurs where the absolute value of the Polyakov loop is unity with all
$\varphi$ fluctuations having a positive mass squared $(R^{-2}+\lambda
T^2)$. 

However, the qualitative picture of the potential changes at 
temperatures $T\gtrsim{(\sqrt\lambda R)}^{-1}$. While the global minimum
continues to be at $\varphi_{ai}=\theta_i=0$, the features of the
effective potential away from the minimum can change quite dramatically.
Expanding the action to quadratic order in $\tilde{\varphi}$ near 
$\tilde{\varphi}\approx 0$, for a generic configuration of
eigenvalues of the Polyakov loop, we find
\\
\SP{
&S_{\rm eff}(\theta,\varphi)
\simeq \\
&2(TR)^3\pi^2 \sum_{ij=1}^N\left(
{R^{-2}\over \lambda T^2}\;\tilde{\varphi}_{aij}^2
\;\;+\;{4\over \pi^2}\;{\tilde{\varphi}_{aij}}^2\sum_{k=1}^\infty 
{1\over
  (2k-1)^2}\cos((2k-1)\theta_{ij})+\right.
\\&\left.
\qquad\qquad\qquad
-{16\over \pi^2} \sum_{k=1}^\infty {1\over (2k-1)^4}
\cos((2k-1)\theta_{ij})+ 
\cdots
\right)\ .
\label{pott2}
}
\\
Clearly, when the radius of the three-sphere becomes comparable to the
Debye screening length, it is conceivable that values of 
$\{\theta_{i}\}$ for which the cosines in the above formula 
turn negative, can lead to instabilities in the $\varphi$ direction. 
 
\begin{figure}[ht]
\centerline{\includegraphics[width=3in]{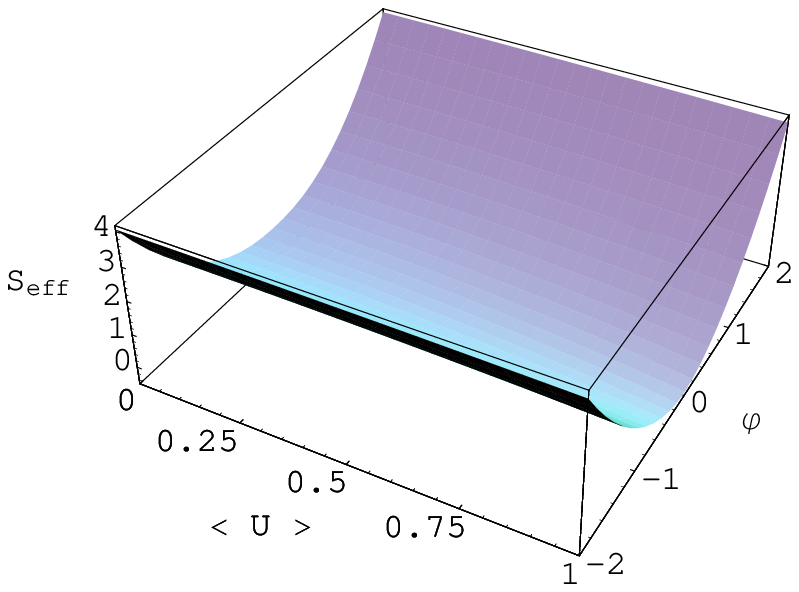}\quad\includegraphics[width=3in]{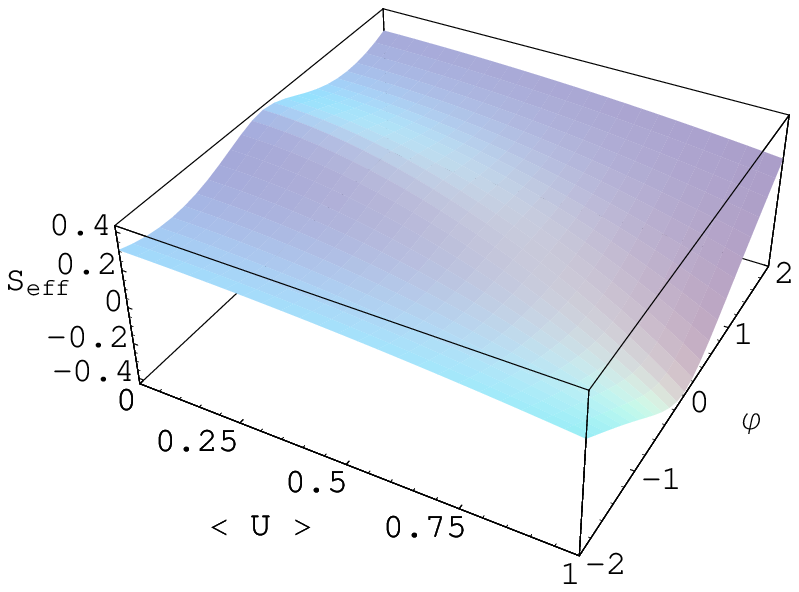}} 
\caption{\footnotesize The effective action is plotted for $T<T_c$
  on the left and $T>T_c$ on the right, as a function of $\varphi$ and
  the Polyakov 
  loop $\langle U\rangle$ for $SU(2)$ gauge group.}
\label{effective}
\end{figure}

Indeed, we will find that there is a critical temperature below which
there are no  
unstable $\varphi$ fluctuations for a given configuration of $\{\theta_i\}$,
\EQ{ T_c =\sqrt{\kappa_N\over\lambda} R^{-1}\ .
\label{hyo}
}
where $\kappa_N$ is an $N$-dependent constant.
At this temperature, an unstable direction develops around the
configuration where the eigenvalues are uniformly distributed with
zero expectation value for the Polyakov loop. Above
the critical temperature, the unstable $\varphi$ mode makes an
appearance for a wider range of values of the Polyakov loop.
\\\\
\underline{\it Example: SU(2)}
\\\\
The situation is easiest to illustrate in 
the $SU(2)$ theory where the set $\{\theta_i\}$ and $\{\varphi_{ai}\}$,
each have only a single independent degree of freedom since
\EQ{\theta_1+\theta_2=2\pi,\qquad\varphi_{a1}+\varphi_{a2}=0.}
The $SU(2)$ Polyakov loop is then
\EQ{U = \cos\theta_1,\qquad\qquad 0\leq\theta_1\leq\pi.}
Rewriting the full $SU(2)$ 
effective potential \eqref{fulleff} 
in terms of the variables $\theta_1$
and $\varphi_1$ we find the behaviour shown 
in Figure \ref{effective}.

The $SU(2)$ theory has a critical temperature
\EQ{T_c = \sqrt {2\over\lambda} R^{-1}.}
For temperatures in the range $R^{-1}\ll T < T_c$, there is a saddle point at 
$\langle U
\rangle=\varphi=0$  
with one unstable direction along the $U$-axis.  This saddle point is
actually  stable below the Hagedorn or deconfinement temperature on 
$\sph{3}$ \cite{Aharony:2003sx} and is the confined phase with zero
expectation value for the Polyakov 
loop. 

The situation changes
beyond the critical temperature $T_c$ when
a second unstable direction appears at $\langle U\rangle =0$. 
The full effective potential \eqref{fulleff} now exhibits new extrema or
saddle points with 
\EQ{\left({\varphi_{a1}\over T}\right)^2 \neq 0}
as in Figure \ref{effective}. It is easy to see that these saddle points must
exist. At $\langle U\rangle =0$, we have seen that there is an
instability at $\varphi_1=0$. For large values of $\varphi_1$, the
quadratic term in \eqref{fulleff} dominates since ${\cal H}(\varphi)$
vanishes exponentially. Hence there must be a point for intermediate
values of $(\varphi_1/T)$ which is stable to fluctuations along this
direction. Importantly, in Figure \ref{effective} we see that these points are
unstable to rolling to the $\langle U\rangle=1$ vacuum. Hence
these new extrema are indeed saddle points.
The
instability to fluctuations in $\varphi$ around $\varphi=0$ persists
for a finite range of values of the Polyakov loop $\langle
U\rangle$. At very high temperatures above $T_c$, 
all points with $\varphi=0$ and $0\leq\langle U\rangle\leq
{1\over\sqrt 2}$ acquire an unstable mode along the $\varphi$ axis. 
\\
\\
\underline{\it $SU(N)$ at large $N$}
\\\\
The above picture generalizes rather directly for $SU(N)$ gauge group.
The unstable extremum at $\varphi_{ai}=0$ is characterized by a
uniform distribution of the eigenvalues of the Polyakov loop,
\EQ{\theta_j= {2\pi\over N}j + c(N)\;,\qquad j=1,2,\ldots N}
where $c$ is a constant chosen so that  
\EQ{\sum_{i=1}^N\theta_i=\;0 \;{\rm mod} \;2\pi.}
It is a stable ground state of the theory
below the Hagedorn temperature on
$\sph{3}$ \cite{Aharony:2003sx} and leads to a vanishing Polyakov loop
and infinite free energy for colored sources. At strong coupling (and
large $N$) and
below the deconfinement transition, this phase is dual to the 
thermal AdS space saddle point of semiclassical Euclidean gravity.

The temperatures we are considering, $T\gtrsim (\sqrt\lambda R)^{-1}$
at weak coupling are far above the deconfinement transition and hence
$\langle U\rangle =0$ is only an unstable extremum. The question 
we are interested in is, whether there are new unstable directions about
this saddle point. This may be answered by examining the quadratic
form in $\varphi$ \eqref{pott2}. The associated matrix is a circulant,
whose eigenvalues may be determined  relatively easily (see Appendix
A). In the large $N$ limit, we find that as the temperature is
increased beyond a critical value new unstable directions
start to appear, with the first negative eigenvalue
occurring at the critical temperature
\EQ{T_c= {\pi\over \sqrt {2\lambda}}\;R^{-1}.\label{critical}}
As the temperature is increased beyond this, we find more negative
modes, each one appearing when the temperature hits the critical
values (Appendix A),
\EQ{T_c^{(l)}= (2l-1){\pi \over \sqrt {2 \lambda}}R^{-1},\qquad
  l=1,2,\ldots.\label{highercrit}}
The number of unstable directions is $(N/2-1)$ when $N$ is even and
$(N-1)/2$ for $N$ odd, and hence is formally infinite in the
large $N$ limit in the one loop approximation. Interestingly, this is
consistent 
with what is observed for the small AdS black hole in classical gravity 
\cite{Hubeny:2002xn}. We should
however, bear in mind that in weakly coupled field theory we expect 
perturbation theory for the constant modes or low lying harmonics on
$S^3$ to  break down at high enough temperatures where the size
of the three sphere approaches the non-perturbative magnetic screening
length $R\sim (\lambda T)^{-1}$. Therefore we can trust
\eqref{highercrit} for $(2l-1)\ll{1\over\sqrt\lambda}$.
We also point out that the modes that we see becoming unstable
are constant on $S^3\times S^1$, the higher Kaluza-Klein harmonics
remain massive at these temperatures and within the regime
of validity of the one loop approximation.

As in the $SU(2)$ example we then find new extrema where 
\EQ{\sum_{a=1}^6 \varphi_{ai}^2 \neq 0\,.}
The existence of these saddle points follows from the instabilities at
$\varphi=0$ 
and the fact that at large $\varphi$  the quadratic mass term
dominates the potential. This can be verified by numerically
plotting the full expression for the potential \eqref{fulleff}. Our
potential is written in terms of the $SO(6)$-invariant radial
coordinate $\sum_a {\varphi_{ai}^2}$ in the space of the six scalars. 
In a non-compact space, a non-zero value for this coordinate at the new
saddle points would be interpreted as 
breaking the global R-symmetry of the
$\N=4$ theory to $SO(5)$
\EQ{SO(6)\longrightarrow SO(5).}
How can we understand this in the context of the field theory on a compact
space, namely the three-sphere, where we do not expect to 
see symmetry breaking.
\\\\
\\\\\
\underline{\it Symmetry breaking at large N}\\
\\
Symmetry breaking is not normally possible in a finite volume since the wave
function spreads over the vacuum manifold or points in field space 
which are related by
the symmetry. This means that we must average
over the orbit of the symmetry group,
 or include in the partition function, an 
integral over such states. Concretely, introducing the notation
\EQ{\rho_i =\left(\sum_{a=1}^6 \varphi_{ai}^2\right)^{1\over
    2}\,\qquad i=1,2,\ldots N}
for the radial mode of scalar eigenvalues, if
$\rho_i$ is allowed a non-vanishing value, then we must integrate over the 
corresponding orbit $SO(6)/SO(5)\simeq S^5$. Formally then, we must
write the partition function as
\EQ{Z=\int \prod_{j=1}^N (d\rho_j \; \rho_j^5 )\;(d\Omega_5)^N \;
e^{-S_{\rm eff}[{\rho_i}]}.}
Actually the above is not completely correct since the orbit of the
global symmetry group is the 
symmetric product space, ${\rm Sym}_N(SO(6)/SO(5))\simeq {\rm Sym}_N
(S^5)$, due to the action of the Weyl subgroup of $SU(N)$. 
However, what is important for us is that the space is
$5N$-dimensional and the full partition function is 
\EQ{Z\sim\int \prod_{j=1}^N (d\rho_j)
\;e^{-S_{\rm eff}[{\rho_i}]+5\sum_{i=1}^N\ln\;\rho_i}.}
However $S_{\rm eff}$ scales as $N^2$ in the large $N$ limit while the
measure from averaging over the orbit of the symmetry group is
of ${\cal O}(N^1)$. Hence, in the large $N$ limit we may neglect the
contribution from the latter to compute the saddle points of the
effective action. Note that taking the volume of the
space to infinity would also have a similar effect since $S_{\rm eff}$
scales extensively with the volume of the three-sphere.

Symmetry breaking may now be defined in the large $N$ theory in the
conventional way. At finite $N$ we introduce a source which breaks the
$SO(6)$ symmetry,
\EQ{\delta S = N^2 \epsilon_i\;\varphi_{ai}.}
Taking $N$ to infinity first, followed by $\epsilon$ to zero 
permits breaking of the $SO(6)$ symmetry.\footnote{
Global (chiral) symmetry breaking in large $N$ gauge theory on a finite
volume space has been studied in a different context in 
\cite{Narayanan:2004cp}}

As in the $SU(2)$ example, the unstable directions persist 
as we move away from $\langle U\rangle =0$, the uniform distribution for the
$\theta_i$. Indeed, as some of the $\theta_i$ start to clump together,
the cosines in \eqref{pott2} cease to be negative and some of the
negative modes start disappearing. This is true in particular
when the distribution develops a gap, an extreme example of which is
the delta function distribution associated to the big AdS black hole
which is the global minimum of the action. 

\subsubsection{The gravity interpretation}

We now compare and contrast the weak coupling, high temperature
picture obtained above of large $N$, $\N=4$ SYM on $\sph{3}\times
\sph{1}$ with the known  
strongly coupled dual, namely gravity/string theory on $\ads{5}\times
\sph{5}$. 

The free ${\cal N}=4$ theory on $\sph{3}$ analyzed in
\cite{Aharony:2003sx} exhibits two different phases at large $N$, 
a confined phase
and a deconfined phase separated by a first order phase
transition at $T =T_H =1/(R \ln(7-4\sqrt 3)^{-1})$. It is 
believed that these two phases at $\lambda =0$ 
are the infinitely curved, stringy
versions of thermal AdS space and the big AdS Schwarzschild black
hole which are separated by a first order Hawking-Page
transition. 

In the free ($\lambda=0$) Yang-Mills theory, above $T_H$ the thermal
AdS extremum   
is unstable. 
Our field theory results above show that 
the weakly interacting large $N$ field theory develops additional
(unstable) saddle
points above a critical temperature
\EQ{T_c={\pi\over \sqrt {2\lambda}}\;R^{-1}.}
These configurations with scalar expectation values only preserve an 
$SO(5)$ subgroup of the $SO(6)$ R-symmetry of $\N=4$ theory. 
We stress that these saddle points are not
themselves stable, but have lower actions than the ``thermal AdS''
configuration (characterized by $\langle U\rangle =\varphi_{ia}=0$)
which develops unstable directions towards these new saddle points at
high 
temperature. 

We interpret these new unstable directions as the infinite curvature
or string scale  
manifestations of the Gregory-Laflamme instability of the small AdS
Schwarzschild black hole, a phenomenon which
is observed at weak curvature or strong 't Hooft coupling
($\lambda\rightarrow\infty$). As 
described in section 2, in semiclassical gravity this instability 
is a dynamical instability at the small AdS black hole
saddle point which is already thermodynamically unstable. The
dynamical instability triggers the localization of the small black hole
on the $S^5$ breaking the $SO(6)$ isometry to $SO(5)$.

There is of course a crucial distinction between the weak and strong
coupling regimes. The dynamical instability we see occurs at
temperatures $T\gtrsim  O(R^{-1}/\sqrt\lambda)$ which is far above
the Hagedorn temperature. 
At these temperatures the saddle point corresponding to the
small black hole has already disappeared \cite{Aharony:2003sx} or
merged with the thermal AdS configuration
\cite{Alvarez-Gaume:2005fv}. Hence 
it is the $\langle U\rangle =0$ extremum, associated to thermal AdS at
strong coupling, which becomes ``unstable to localization on
$\sph{5}$''.  

It is easy to see how the the strong and weak coupling pictures might
match on smoothly.
As the 't Hooft coupling is increased, the critical temperature
\eqref{critical} decreases at weak coupling. It is conceivable that as
we move towards strong coupling the critical temperature $T_c$ at which the
new instabilities are triggered, eventually becomes comparable to and
lower than the 
Hagedorn temperature where the small black hole exists as a saddle
point of the action. Hence for temperatures in the range $T_c< T <
T_H$, the small black hole saddle point can be expected to be unstable
to new saddle points which break the $SO(6)_R$ down to $SO(5)_R$. As
we have already mentioned in the introduction, there is also the
possibility that at relatively small 't Hooft couplings with 
dual gravity curvatures approaching the string
scale, the small black hole likely makes a transition to a highly excited
state of strings before any  Gregory-Laflamme like instabilities can
kick in.

\section{Summary and conclusions}

In this paper we have performed an explicit computation of a joint one
loop effective action for the eigenvalues of constant scalar
fields and the 
Polyakov loop in $SU(N)$ $\N=4$ SYM on $\sph{3}\times \sph{1}$.
Our perturbative computation is valid for temperatures 
$0\leq T\ll {1\over \lambda}R^{-1}$ and in particular for temperatures
where the size
of the three sphere approaches the Debye screening length of the
Yang-Mills plasma $R \sim (\sqrt\lambda T)^{-1}$.

We find at large $N$, above a critical temperature
\EQ{T_c={\pi\over\sqrt{2 \lambda}}\;R^{-1}}
the effective potential exhibits new extrema with non-zero values for
the $\N=4$ scalar fields. The emergence of these saddle points
is accompanied by new unstable directions at the thermal AdS
extremum which was characterized by a uniform distribution of Polyakov
loop eigenvalues. 
The latter is already unstable at temperatures below $T_c$
but new instabilities along the scalar directions appear at $T_c$.
We believe these unstable directions to be the small 't Hooft  coupling
manifestations of a well known dynamical instability in the gravity
dual of the strongly coupled gauge theory -- the Gregory-Laflamme
localization instability of the small black hole in $\ads{5}\times\sph{5}$. 
Note that the weak coupling instability we have found is not associated to a
small black hole saddle point since the latter does not exist at $T_c$
or has
merged with the thermal AdS saddle point at the Hagedorn temperature 
far below $T_c$.

Many open questions remain, chief among these being the physical
meaning of the instabilities of the thermal AdS saddle
point at these high temperatures, and how they turn into dynamical
instabilities of the small black hole at intermediate or strong
coupling. Possible answers to this question have been presented in
the paper and are summarized in a phase plot in Figure \ref{phase}.
 
We also find that whenever the temperature reaches a critical value
$T_c^{(l)}=(2l-1)T_c$ for $l=1,2,3,\ldots$ a new negative mode shows
up. A similar infinite set of negative modes is also expected from the 
classical 
gravity analysis of \cite{Hubeny:2002xn}. It would be interesting to
understand the similarities between the two pictures better,
particularly whether the linear dependence of $T_c^{(l)}$ on $l$ can be
understood at all from the gravity perspective, even though the latter is
only valid at strong 't Hooft coupling. It is somewhat
tantalizing that the values of $T_c^{(l)}$ for the first five negative
modes obtained numerically in \cite{Hubeny:2002xn}
\footnote{The authors of \cite{Hubeny:2002xn} give the values of $r_+$
for each of these modes, which can be easily converted to a
temperature.} appear to exhibit
this linear behaviour.

A natural extension of the results of this paper is to include a
finite chemical potential 
for an $SO(2)$ subgroup of the $SO(6)$ R-symmetry. The corresponding
gravity dual involves charged black holes. These also exhibit
Gregory-Laflamme instabilities and we explore them at weak coupling
in   the holographic dual in \cite{nextpaper}. The weak coupling
analysis appears to exhibit stable saddle points with global symmetry
breaking at finite chemical potential.  
The phase structure for
R-charged black holes was discussed in
\cite{Chamblin:1999tk,Cvetic:1999ne,Basu:2005pj} and the dual field
theory phase diagram at weak coupling has been explored recently in
\cite{Yamada:2006rx}. 
  
The effective potential we have computed is a static
effective potential since we calculate it in the Euclidean theory on
$\sph{3}\times\sph{1}$. It is difficult to interpret it as an
effective potential in the Lorentzian field theory on $\sph{3}\times
{\mathbb R}$ since it is not clear what the Polyakov loops mean in the
Lorentzian theory. However, there is a different analytic continuation
to Lorentzian signatures where our effective potential may be usefully
interpreted. This involves analytically continuing one of the angular
directions on the three-sphere to Lorentzian signature which maps
$\sph{3}\times \sph{1}$ to ${\rm dS}_3\times \sph{1}$, leaving the
thermal circle as a spatial circle. As discussed originally in
\cite{Balasubramanian:2005bg} this is the
boundary of a ``topological black hole'' constructed as a certain
orbifold of $\ads{5}$ \cite{Banados:1998dc} . 
The topological black hole decays to an AdS
bubble of nothing 
\cite{Balasubramanian:2002am} where the bounce solution is the Euclidean
small AdS black hole which is unstable to localization on the $\sph{5}$.
As pointed out by the authors of \cite{Balasubramanian:2005bg}, this
should be described in a dual field theory via an effective potential
that allows rolling in a direction breaking the $SO(6)$
symmetry. The methods presented in our paper and our effective
potential naturally allow such a  holographic description of the
process of 
gravitational decay via the bubble of nothing, at least in an adiabatic
approximation. 
\\\\
{\bf{Acknowledgments}}\\\\
We would like to thank Gert Aarts, Sean Hartnoll and Shiraz
Minwalla for enjoyable discussions and very useful 
comments. SPK and AN are supported by PPARC Advanced Fellowships.

\startappendix
\Appendix {Unstable directions of the effective potential}

In this appendix, we perform the small fluctuation analysis of the effective
potential expanded about $\varphi\approx 0$ for a uniformly
distributed configuration of Polyakov
loop eigenvalues \eqref{pott2}. This will tell us about instabilities
around the thermal AdS saddle point as a function of temperature.

We are interested in the eigenvalues
of the quadratic form in $\tilde{\varphi_i}\tilde{\varphi_j}$ appearing in
\eqref{pott2} which can be rewritten as (neglecting overall
multiplicative constants)
\\
\SP{\sum_{ij=1}^N\tilde{\varphi_i}\tilde{\varphi_j} \Big(
&\delta_{ij}\left[{R^{-2}\over\lambda T^2} \;(N-1)-{1\over 2}
+{1\over 2N} \delta_{N, {\rm odd}}\right]-\\
&(1-\delta_{ij})\left[{R^{-2}\over\lambda T^2} +{4\over \pi^2}
\sum_{k=1}^\infty {1\over (2k-1)^2} \cos\left( {2\pi\over
    N}(2k-1)(i-j)\right)\right] 
\Big), \label{quadra}}
\\
where we have used $\theta_{ij}= 2\pi(i-j)/N$ at the thermal AdS
saddle point.

The quadratic form above is an example of an $N\times N$ circulant matrix {\it
  i.e.} a matrix each of whose rows (or columns) can be 
  obtained by cyclic permutations of the elements of one particular
  row (or column). The $N$ eigenvalues of a circulant matrix with a
  row of the form $(x_1,x_2,\ldots x_N)$ are given by 
\EQ{\lambda_\ell=\sum_{m=1}^{N} x_m \; e^{{2\pi i \over
  N}(m-1)\ell}\qquad\ell= 0,1,2,\ldots N-1.}

Applying this formula to the quadratic form \eqref{quadra}, we can find
the eigenvalues easily. The expressions simplify at large $N$ and we
quote the results below 
\SP{&\lambda_0=0\\\\
&\lambda_{2\ell-1}= N\left[\left({R^{-2}\over\lambda T^2}\right)
  -{2\over 
    \pi^2}{1\over (2\ell-1)^2} + O\left({1\over N^2}\right)\right]\\\\
&\lambda_{2\ell}=N\left[ \left({R^{-2}\over\lambda T^2}\right) + O({1 \over N^2}) \right]}
with $\ell=1,2,\ldots$.

It is clear that each of the eigenvalues with odd labels becomes
negative at a critical temperature
\EQ{
T_{\ell}= {(2\ell-1)\pi\over \sqrt {2 \lambda}} R^{-1}.}
and the first instability occurs at 
\EQ{T_c={\pi\over \sqrt {2 \lambda}} R^{-1}.}

\end{document}